\documentclass{article}
\def\~{\tilde}
\usepackage{graphicx}

\begin{document}

\title{Consistent histories and the Bohm approach.}
\author{B. J. Hiley and O. J. E. Maroney\\
Theoretical Physics Research Unit\\
Birkbeck College, University of London\\
Malet Street, London WC1E 7HX, England}
\date{13 September 2000}
\maketitle

\begin{abstract}
In a recent paper Griffiths claims that the consistent histories
interpretation of quantum  mechanics gives rise to results that contradict
those obtained from the Bohm interpretation. This is in spite of the fact
that both claim to provide a realist interpretation of the formalism without
the need to add any new {\it mathematical} content and both always produce
exactly the same probability predictions of the outcome of experiments. In
contrasting the differences Griffiths argues that the consistent histories
interpretation provides a more physically reasonable account of quantum
phenomena. We examine this claim and show that the consistent histories
approach is not without its difficulties.
\end{abstract}

\section{Introduction}

It is well known that realist interpretations of the quantum formalism are
known to be notoriously difficult to sustain and it is only natural that the
two competing approaches, the consistent history interpretation (CH) \cite
{RG84} \cite{RG96} and the Bohm interpretation (BI)\cite{BH87}\cite{BH93},
should be carefully compared and contrasted. Griffiths \cite{RG99} is right
to explore how the two approaches apply to interferometers of the type shown
in figure 1.

Although the predictions of experimental outcomes expressed in terms of
probabilities are identical, Griffiths argues that, nevertheless, the two
approaches actually give very different accounts of how a particle is
supposed to pass through such an interferometer. After a detailed analysis
of experiments based on figure 1, he concludes that the CH approach gives a
behaviour that is `physically acceptable', whereas the Bohm trajectories
behave in a way that appears counter-intuitive and therefore `unacceptable'.
This behaviour has even been called `surrealistic' by some authors\footnote{%
This original criticism was made by Englert et al. \cite{ESSW}. An extensive
discussion of this position has been presented by Hiley, Callaghan and
Maroney \cite{HCM}.}. Griffiths concludes that a particle is unlikely to
actually behave in such a way so that one can conclude that the CH
interpretation gives a `more acceptable' account of quantum phenomena.

\begin{figure}[t]
\includegraphics{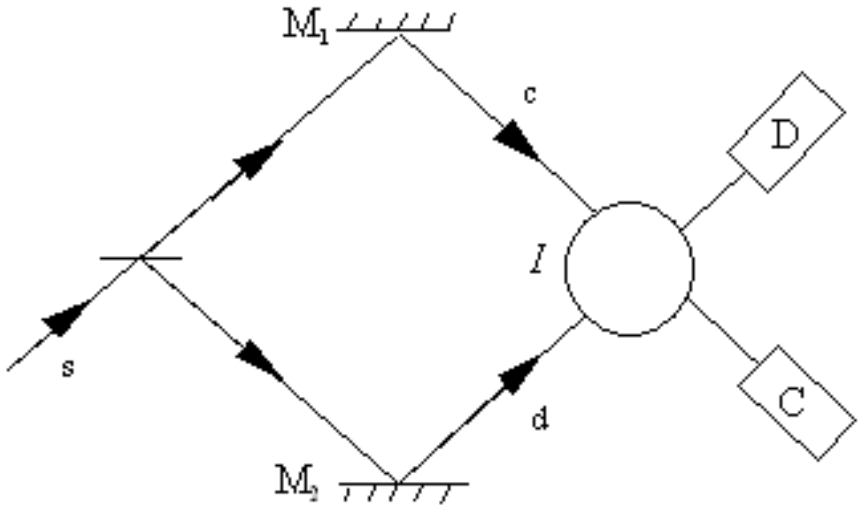}
\caption{Simple interferometer}
\end{figure}

Notice that these claims are being made in spite of the fact no new
mathematical structure whatsoever is added to the quantum formalism in
either CH or BI, and in consequence all the experimental predictions of both
CH and BI are identical to those obtained from standard quantum mechanics.
Clearly there is a problem here and the purpose of our paper is to explore
how this difference arises. We will show that CH is not without its
difficulties.

We should remark here in passing that these difficulties have already been
brought out be Bassi and Ghirardi \cite{BGH} \cite{BGH1} \cite{BGH2} and an
answer has been given by Griffiths \cite{RG20}. At this stage we will not
take sides  in this general debate. Instead will examine carefully how the
analysis of the particle behaviour in CH when applied to the interferometer
shown in figure 1 leads to difficulties similar to those highlighted by
Bassi and Ghirardi \cite{BGH}.

\section{Histories and trajectories}

The first problem we face in comparing the two approaches is that BI uses a
mathematically well defined concept of a trajectory, whereas CH does not use
such a notion, defining a more general notion of a history.

Let us first deal with the Bohm trajectory, which arises in the following
way. If the particle satisfies the Schr\"{o}dinger equation then the
trajectories are identified with the one-parameter solutions of the real
part of the Schr\"{o}dinger equation obtained under polar decomposition of
the wave function \cite{BH93}. Clearly these one-parameter curves are
mathematically well defined and unambiguous.

CH does not use the notion of a trajectory. It uses instead the concept of a
history, which, again, is mathematically well defined to be a series of
projection operators linked by Schr\"{o}dinger evolution and satisfying a
certainty consistency condition \cite{RG84}. Although in general a history
is not a trajectory, in the particular example considered by Griffiths,
certain histories can be considered to provide approximate trajectories. For
example, when particles are described by narrow wave packets, the history
can be regarded as defining a kind of broad `trajectory' or `channel'. It is
assumed that in the experiment shown in figure 1, this channel is narrow
enough to allow comparison with the Bohm trajectories.

To bring out the apparent difference in the predictions of the two
approaches, consider the interferometer shown in figure 1. According to CH
if we choose the correct framework, we can say that if $C$ fires, the
particle must have travelled along the path $c$ to the detector and any
other path is regarded as ``dynamically impossible" because it violates the
consistency conditions. The type of trajectories that would be acceptable
from this point of view are sketched in figure 2. In contrast a pair of
typical Bohm trajectories \footnote{%
Detailed examples of these trajectories will be found in Hiley, Callaghan
and Maroney \cite{HCM}.} are shown in figure 3 . Such trajectories are
clearly not what we would expect from our experience in the classical world.
Furthermore there appears, at least at first sight, to be no visible
structure present that would `cause' the trajectories to be `reflected' in
the region $I$, although in this region interference between the two beams
is taking place.


\begin{figure}[t]
\includegraphics{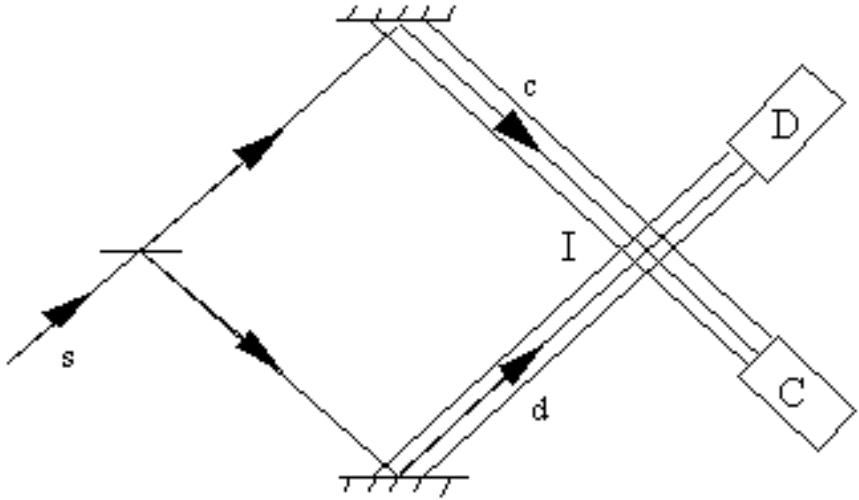}
\caption{The CH `trajectories'.}
\end{figure}

In the Bohm approach, an additional potential, the quantum potential,
appears in the region of interference and it is this potential that has a
structure which `reflects' the trajectories as shown in figure 3. (See Hiley
et al. \cite{HCM} for more details).

\begin{figure}[t]
\includegraphics{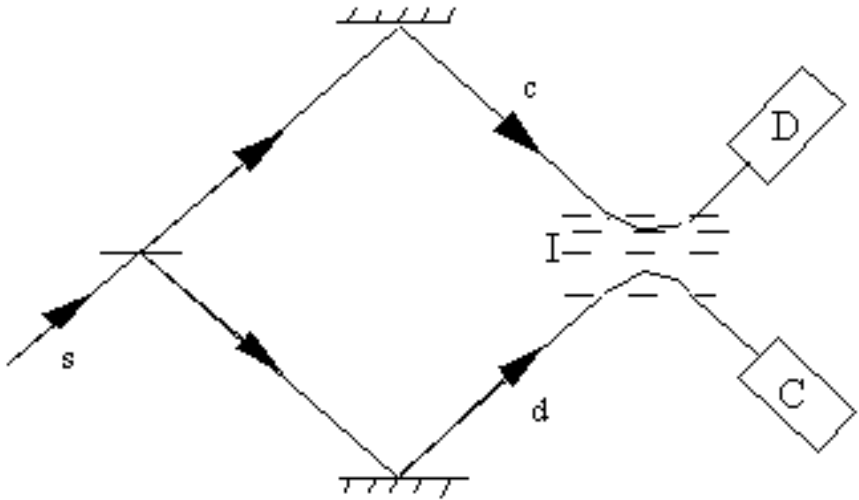}
\caption{The Bohm trajectories.}
\end{figure}

In this short note we will show that the conclusions reached by Griffiths 
\cite{RG99} cannot be sustained and that it is not possible to conclude that
the Bohm `trajectories' must be `unreliable' or `wrong'. We will show that
CH cannot be used in this way and the conclusions drawn by Griffiths are not
sound.

\section{The interference experiment}

Let us analyse the experimental situation shown in figure 1 from the point
of view of CH.  A unitary transformation $U(t_{j+1}, t_{j})$ is used to
connect set of projection operators at various times. The times of interest
in this example will be $t_{0}, t_{1}$, and $t_{2}$. $t_{0}$ is a time
before the particle enters the beam splitter, $t_{2}$ is the time at which a
response occurs in one of the detectors $C$ or $D$ and $t_{1}$ is some
intermediary time when the particle is in the interferometer before the
region I is reached by the wave packets.

The transformation for $t_{0}\rightarrow t_{1}$ is 
\begin{equation}
|\psi_{0}\rangle = |sCD\rangle_{0}\rightarrow \frac{1}{\surd{2}}[%
|cCD\rangle_{1} + |dCD\rangle _{1}]
\end{equation}

The transformation for $t_{1}\rightarrow t_{2}$ is, according to Griffiths 
\cite{RG93},\cite{RG99} 
\begin{equation}
|cCD\rangle_{1} \rightarrow |C^{*}D\rangle_{2},\hspace{0.5in} \mbox{and} %
\hspace{0.5in} |dCD\rangle_{1}\rightarrow |CD^{*}\rangle_{2}
\end{equation}

These lead to the histories 
\begin{equation}
\psi_{0} \otimes c_{1} \otimes C_{2}^{*},\hspace{0.5in} \mbox{and} %
\hspace{0.5in} \psi_{0} \otimes d_{1}\otimes D_{2}^{*)}
\end{equation}
Here $\psi_{0}$ is short hand for the projection operator $%
|\psi\rangle\langle \psi|$ at time $t_{0}$ etc.

These are not the only  possible consistent histories but only these two
histories are used by Griffiths to make judgements about the Bohm
trajectories. The two other possible histories 
\begin{equation}
\psi_{0} \otimes d_{1} \otimes C_{2}^{*}, \hspace{0.5in} \mbox{and} %
\hspace{0.5in} \psi_{0}\otimes c_{1} \otimes D_{2}^{*}
\end{equation}
have zero weight and are therefore deemed to be {\em dynamically impossible}.

The significance of the histories described by equation (3) is that they
give rise to new conditional  probabilities that {\em cannot} be obtained
from the Born probability rule \cite{RG98}. These conditional probabilities
are 
\begin{equation}
Pr(c_{1}|\psi_{0}\wedge C_{2}^{*}) = 1,\hspace{0.5in} Pr(d_{1}|\psi_{0}%
\wedge D_{2}^{*}) = 1.
\end{equation}

Starting from a given initial state, $\psi_{0}$, these probabilities are
interpreted as asserting that when the detector $C$ is triggered at $t_{2}$,
one can be certain that, at the time $t_{1}$, the particle was in the
channel {\em c} and not in the channel {\em d}. In other words when $C$
fires we know that the triggering particle must have travelled down path $c$
with certainty.

{\em This is the key new result from which the difference between the
predictions of CH and the Bohm approach arises}. Furthermore it must be
stressed that this result cannot be obtained from the Born probability rule
and is claimed by Griffiths \cite{RG98} to be a new result that does not
appear in standard quantum theory\footnote{%
It should be noted that the converse of (5) must also hold. Namely, if $C$
does not fire then we can conclude that at $t_{1}$ the particle was not in
pathway $c$. In other words $Pr(c_{1}|\psi_{0}\wedge C_{2}) = 0$}.

Looking again at figure 1, we notice that there is a region $I$ where the
wave packets travelling down $c$ and $d$ overlap. Here interference can and
does take place. In fact fringes will appear along any vertical plane in
this region as can be easily demonstrated. Indeed this interference is
exactly the same as that produced in a two-slit experiment. The only change
is that the two slits have been replaced by two mirrors. Once this  is
realised alarm-bells should ring because the probabilities in (5) imply that
we know with certainty through which slit the particle passed. Indeed
equation (5) shows that the particles passing through the lower slit will
arrive in the upper region of the fringe pattern, while those passing
through the upper slit will arrive in the lower half \footnote{%
Notice that in criticising the Bohm approach, it is this consistent history
interpreted as a `particle trajectory' that is contrasted with the Bohm
trajectory. The Bohm approach reaches the opposite conclusion, namely, the
particle that goes through the top slit stays in the top part of the
interference pattern \cite{PDH}}.

Recall that Griffiths claims CH provides a clear and consistent account of
standard quantum mechanics, but the standard theory denies the possibility
of knowing which path the particle took when interference is present. Thus
the interpretation of equation (5) leads to a result that is not part of the
standard quantum theory and in fact contradicts it. Nevertheless CH uses the
authority of the standard approach to strengthen its case against the Bohm
approach. Surely this cannot be correct.

Indeed Griffiths has already discussed the two-slit experiment in an earlier
paper \cite{RG94}. Here he argues that CH does not allow us to infer through
which slit the particle passes. He writes; -

\begin{quote}
Given this choice at $t_{3}$ [whether $C$ or $D$ fires], it is inconsistent
to specify a decomposition at time $t_{2}$ [our $t_{1}$] which specifies
which slit the particle has passed through, i.e., by including the projector
corresponding to the particle being in the region of space just behind the $A
$ slit [our $c$], and in another region just behind the $B$ slit [our $d$].
That is (15) [the consistency condition] will not be satisfied if projectors
of this type at time $t_{2}$ [our $t_{1}$] are used along with those
mentioned earlier for time $t_{3}$.
\end{quote}

The only essential difference between the two-slit experiment and the
interferometer described by equation (3)  above is in the position of the
detectors. But according to CH measurement merely reveals what is already
there, so that the position of the detector in the region $I$ or beyond
should not affect anything. Thus there appears to be a contradiction here.

To emphasise this difficulty we will spell out the contradiction again. The
interferometer in figure 1 requires the amplitude of the incident beam to be
split into two before the beams are brought back together again to overlap
in the region $I$. This is exactly the same process occurring in the
two-slit experiment. Yet in the two-slit experiment we are  not allowed to
infer through which slit the particle passed while retaining interference,
whereas according to Griffiths we are allowed to talk about which mirror the
particle is reflected off, presumably without also destroying the
interference in the region $I$. We will return to this specific point again
later.

One way of avoiding this contradiction is to assume the following: -

1. If we place our detectors in the arms $c$ and $d$ before the interference
region $I$ is reached then we have the consistent histories described in
equation (3). Particles travelling down $c$ will fire $C$, while those
travelling down $d$ will fire $D$. In this case we have an exact agreement
with the Bohm trajectories.

2. If we place our detectors in the region of interference $I$ then,
according to Griffiths \cite{RG94}, the histories described by equation (3)
are no longer consistent. In this case CH can say nothing about trajectories.

3. If we place our detectors in the positions shown in figure 1, then,
according to Griffiths \cite{RG99}, the consistent histories are described
by equation (3) again. Here the conditional probabilities imply that all the
particles travelling down $c$ will always fire $C$. Bohm trajectories
contradict this result and show that some of these particles will cause $D$
to fire . These trajectories are shown in figure 3.

It could be argued that this patchwork would violate the {\em one-framework
rule}. Namely that one must either use the consistent histories described by
equation (3) or use a set of consistent histories that do not allow us to
infer off which mirror the particle was reflected. This latter would allow
us to account for the interference effects that must appear in the region $I$%
.

A typical set of consistent histories that do not allow us to infer through
which slit the particle passed can be constructed in the following way.

Introduce a new set of projection operators $|(c + d)\rangle \langle (c + d)|
$ at $t_{3}$ where $t_{1} < t_{3} < t_{2}$. Then we have the following
possible histories 
\begin{equation}
\psi_{0} \otimes (c + d)_{3} \otimes C_{2}^{*},\hspace{0.5in} \mbox {and} %
\hspace{0.5in} \psi_{0} \otimes (c + d)_{3} \otimes D_{2}^{*}
\end{equation}
Clearly from this set of histories we cannot infer any generalised notion of
a trajectory so that we cannot say from which mirror the particle is
reflected. What this means then is that if we want to talk about
trajectories we must, according to CH, use the  histories described by
equation (3) to cover the whole region as, in fact, Griffiths \cite{RG99}
actually does. But then surely the nodes in the interference pattern at $I$
will cause a problem.

To bring out this problem let us first forget about theory and consider what
actually happens experimentally as we move the  detector $C$ along a
straight line towards the mirror $M_{1}$. The detection rate will be
constant as we move it towards the region $I$. Once it enters this region,
we will find that its counting rate varies and will go through several zeros
corresponding to the nodes in the interference pattern. Here we will assume
that the detector is small enough to register these nodes.

Let us examine what happens to the conditional probabilities as the detector
crosses the interference region. Initially according to (5), the first
history gives the conditional probability $Pr(c_{1}|\psi_{0}\wedge
C_{3}^{*}) = 1$. However, at the nodes this conditional probability cannot
even be defined as $Pr(C_{3}^{*}) = 0$. Let us start again with the closely
related conditional probability, derived from the same history $Pr(
C_{3}^{*}|\psi_{0}\wedge c_{1}) = 1$. Now this probability clearly cannot be
continued across the interference region because $Pr(C_{3}^{*}) = 0 $ at the
nodes, while $Pr(\psi_{0}\wedge c_{1}) = 0.5$ regardless of where the
detector is placed. In fact, there is no consistent history that includes
both $c_{1}$ and $C_{3}^{*}$, when the detector is in the interference
region. We are thus forced to consider different consistent histories in
different regions as we discussed above.

If we follow this prescription then when the detector $C$ is placed on the
mirror side of path $c$, before the beams cross at $I$, we can talk about
trajectories and as stated above these trajectories agree with the
corresponding Bohm trajectories. When $C$ is moved right through and beyond
the region $I$, we can again talk about trajectories. However in the
intermediate region  CH does not allow us to talk about trajectories. This
means that we have no continuity across the region of interference and this
lack of continuity means that it is not possible to conclude that any
`trajectory' defined by $\psi_{0} \otimes c_{1} \otimes C^{*}$ before $C$
reaches the interference region is the same `trajectory' defined by the same
expression after $C$ has passed through the interference region. In other
words we cannot conclude that any particle travelling down $c$ will continue
to travel in the same direction through the region of interference and
emerge still travelling in the same direction to trigger detector $C$.

What this means is that CH cannot be used to draw any conclusions on the
validity or otherwise of the Bohm trajectories. These latter trajectories
are continuous throughout {\em all} regions. They are straight lines from
the mirror until they reach the region $I$. They continue into the region of
interference, but no longer travel in straight lines parallel to the initial
their paths. They show `kinks' that are characteristic of interference-type
bunching that is needed to account for the interference \cite{PDH}. This
bunching has the effect of changing the direction of the paths in such a way
that some of them eventually end up travelling in straight lines towards
detector $D$ and not $C$ as Griffiths would like them to do.

Indeed it is clear that the existence of the interference pattern means that
any theory giving relevance to particle trajectories must give trajectories
that do not move in straight lines directly through the region $I$. The
particles must avoid the nodes in the interference pattern. CH offers us no
reason why the trajectories on the mirror side of $I$ should continue in the
same general direction towards $C$ on the other side of $I$. In order to
match up trajectories we have to make some assumption of how the particles
cross the region of interference. One cannot simply use classical intuition
to help us through this region because classical intuition will not give
interference fringes. Therefore we cannot conclude that the particles
following the trajectories before they enter the region $I$ are the same
particles that follow the trajectories after they have emerged from that
region. This requires a knowledge of how the particles cross the region $I$,
a knowledge that is not supplied by CH.

Where the consistent histories (3) could provide a complete description is
when the coherence between the two paths is destroyed. This could happen if
a measurement involving some irreversible process was made in one of the
beams. This would ensure that there was no interference occurring in the
region $I$. In this case the trajectories would go straight through. This
would mean that the conditional probabilities given in equation (5) would
always be satisfied.

But in such a situation the Bohm trajectories would also go straight
through. The particles coming from Mirror $M_{1}$ would trigger the detector 
$C$ no matter where it was placed. The reason for this behaviour in this
case is because the wave function is no longer $\psi_{c} + \psi_{d}$, but we
have two incoherent beams, one described by $\psi_{c}$ and the other by $%
\psi_{d}$. This gives rise to a different quantum potential which does not
cause the particles to be `reflected' in the region $I$. So here there is no
disagreements with CH.

\section{Conclusion}

When coherence between the two beams is destroyed it is possible to make
meaningful inferences about trajectories in CH. These trajectories imply
that any particle reflected from the mirror $M_{1}$ must end up in detector $%
C$. In the Bohm approach exactly the same conclusion is reached so that
where the two approaches can be compared they predict exactly the same
results.

When the coherence between the two beams is preserved then CH must use the
consistent histories described by equation (6). These histories do not allow
any inferences about trajectories to be drawn. Although the consistent
histories described by equation (3) enable us to make inferences about
particle trajectories because, as we have shown they lead to disagreement
with experiment. Unlike the situation in CH the Bohm approach can define the
notion of a trajectory which is calculated from the real part of the
Schr\"{o}dinger equation under polar decomposition. These trajectories are
well defined and continuous throughout the experiment including the region
of interference. Since CH cannot make any meaningful statements about
trajectories in this case it cannot be used to draw any significant
conclusions concerning the validity or otherwise of the Bohm trajectories.
Thus the claim by Griffiths \cite{RG99}, namely, that the CH gives a more
reasonable account of the behaviour of particle trajectories interference
experiment shown in Figure 1 than that provided by the Bohm approach cannot
be sustained.

\end{document}